\documentclass[8.5pt,twoside,twocolumn]{article}
\usepackage[super,sort&compress,comma]{natbib} 
\usepackage{mhchem}
\usepackage{times,mathptmx}
\usepackage{sectsty}
\usepackage{balance} 

\usepackage{graphicx} %eps figures can be used instead
\usepackage{lastpage}
\usepackage[format=plain,justification=raggedright,singlelinecheck=false,font=small,labelfont=bf,labelsep=space]{caption} 
\usepackage{fancyhdr}
\usepackage{color}
\usepackage[normalem]{ulem}

\usepackage{amssymb}

\begin{document}

\thispagestyle{plain}
\fancypagestyle{plain}{
%\fancyhead[L]{\includegraphics[height=8pt]{LH}}
%\fancyhead[C]{\hspace{-1cm}\includegraphics[height=20pt]{CH}}
%\fancyhead[R]{\includegraphics[height=10pt]{RH}\vspace{-0.2cm}}
\renewcommand{\headrulewidth}{1pt}}
\renewcommand{\thefootnote}{\fnsymbol{footnote}}
\renewcommand\footnoterule{\vspace*{1pt}% 
\hrule width 3.4in height 0.4pt \vspace*{5pt}} 
\setcounter{secnumdepth}{5}

\makeatletter 
\def\subsubsection{\@startsection{subsubsection}{3}{10pt}{-1.25ex plus -1ex minus -.1ex}{0ex plus 0ex}{\normalsize\bf}} 
\def\paragraph{\@startsection{paragraph}{4}{10pt}{-1.25ex plus -1ex minus -.1ex}{0ex plus 0ex}{\normalsize\textit}} 
\renewcommand\@biblabel[1]{#1}            
\renewcommand\@makefntext[1]% 
{\noindent\makebox[0pt][r]{\@thefnmark\,}#1}
\makeatother 
\renewcommand{\figurename}{\small{Fig.}~}
\sectionfont{\large}
\subsectionfont{\normalsize} 

\fancyfoot{}
%\fancyfoot[LO,RE]{\vspace{-7pt}\includegraphics[height=9pt]{LF}}
%\fancyfoot[CO]{\vspace{-7.2pt}\hspace{12.2cm}\includegraphics{RF}}
%\fancyfoot[CE]{\vspace{-7.5pt}\hspace{-13.5cm}\includegraphics{RF}}
\fancyfoot[RO]{\footnotesize{\sffamily{1--\pageref{LastPage} ~\textbar  \hspace{2pt}\thepage}}}
\fancyfoot[LE]{\footnotesize{\sffamily{\thepage~\textbar\hspace{3.45cm} 1--\pageref{LastPage}}}}
\fancyhead{}
\renewcommand{\headrulewidth}{1pt} 
\renewcommand{\footrulewidth}{1pt}
\setlength{\arrayrulewidth}{1pt}
\setlength{\columnsep}{6.5mm}
\setlength\bibsep{1pt}

\twocolumn[
  \begin{@twocolumnfalse}
\noindent\LARGE{\textbf{Acoustic measurement of a granular density of modes}}
\vspace{0.6cm}

\noindent\large{\textbf{Eli T. Owens and
Karen E. Daniels$^{\ast}$}}\vspace{0.5cm}
%Please note that \ast indicates the corresponding author(s) but no footnote text is required. 

%\noindent\textit{\small{\textbf{Received Xth XXXXXXXXXX 20XX, Accepted Xth XXXXXXXXX 20XX\newline
%First published on the web Xth XXXXXXXXXX 200X}}}

%\noindent \textbf{\small{DOI: 10.1039/b000000x}}
\vspace{0.6cm}
%Please do not change this text.

\noindent \normalsize{In glasses and other disordered materials, measurements of the vibrational density of states reveal that an excess number of long-wavelength (low-frequency) modes, as compared to the Debye scaling seen in crystalline materials, is associated with a loss of mechanical rigidity. In this paper, we present a novel technique for measuring the density of modes (DOM) in a real granular material, in which we mimic thermal excitations using white noise acoustic waves. The resulting vibrations are detected with piezoelectric sensors embedded inside a subset of the particles, from which we are able to compute the DOM via the spectrum of the velocity autocorrelation function, a technique previously applied in thermal systems. The velocity distribution for individual particles is observed to be Gaussian, but the ensemble distribution is non-Gaussian due to varying widths of the individual distributions. In spite of this deviation from a true thermal system, we find that the DOM exhibits several thermal-like features, including Debye scaling in a compressed hexagonally ordered packing, and an increase in low-frequency modes as the confining pressure is decreased. In disordered packings, we find that a characteristic frequency $f_c$ increases with pressure, but more weakly than has been observed in simulations of frictionless packings.
}
\vspace{0.5cm}
 \end{@twocolumnfalse}
  ]

%Footnotes
%\footnotetext{\dag~Electronic Supplementary Information (ESI) available: [details of any supplementary information available should be included here]. See DOI: 10.1039/b000000x/}

%Please use \dag to cite the ESI in the main text of the article.
%If you article does not have ESI please remove the \dag symbol from the title and the above footnotetext.

\footnotetext{Dept. of Physics, North Carolina State University, Raleigh, NC, USA. E-mail: kdaniel@ncsu.edu}

%\footnotetext{\ddag~Additional footnotes to the title and authors can be included \emph{e.g.}\ `Present address:' or `These authors contributed equally to this work' as above using the symbols: \ddag, \textsection, and \P. Please place the appropriate symbol next to the author's name and include a \texttt{\textbackslash footnotetext} entry in the correct place in the list.}

\section{Introduction}

Simulations of idealized granular materials indicate that they undergo a jamming transition whereby the material becomes rigid and able to support a finite pressure\cite{VanHecke2010,Liu2010}. This transition occurs in systems for which the average coordination number $z$ has increased to the critical value $z_c$, thereby constraining the motion of the particles. Just above the transition, it has been observed in simulations of both frictionless\cite{OHern2003, Silbert2005, Wyart2005, Silbert2009} and frictional\cite{Somfai2007,Henkes2010} granular packings that the density of states $D(\omega)$ exhibits an excess number of low-frequency modes as compared to Debye scaling. Beyond some frequency $\omega^*$, the density of states is observed to deviate from the Debye scaling $D(\omega) \propto \omega^{d-1}$,  where $d$ is the dimensionality of the system. In frictional simulations\cite{Somfai2007,Henkes2010}, this crossover frequency is observed to scale as $\omega^* \propto (z-z_c)$. The spatial eigenmodes associated with frequencies below $\omega^*$ are said to be soft, exhibiting long-wavelength rearrangements.

While real granular materials have no thermal vibrations and the density of states is therefore not strictly defined, it has been possible to observe similar spatial modes via particle tracking and the construction of a covariance matrix\cite{Brito2010,Henkes2012}. Such techniques are similar to those used to measure the density of states for colloids\cite{Ghosh2010a, Chen2010}, where thermal fluctuations of the particles are naturally present. In experiments on an oscillated granular material, the soft modes take the form of collective rearrangements which occur as the system unjams\cite{Brito2010}, and the number of these long-wavelength modes increases as the packing fraction approaches the jamming transition. A disadvantage of using the covariance matrix method is that it requires visual access to particles, and can therefore only be used in highly idealized granular systems.

\begin{figure*}
\centering
  \includegraphics[width=.8\linewidth]{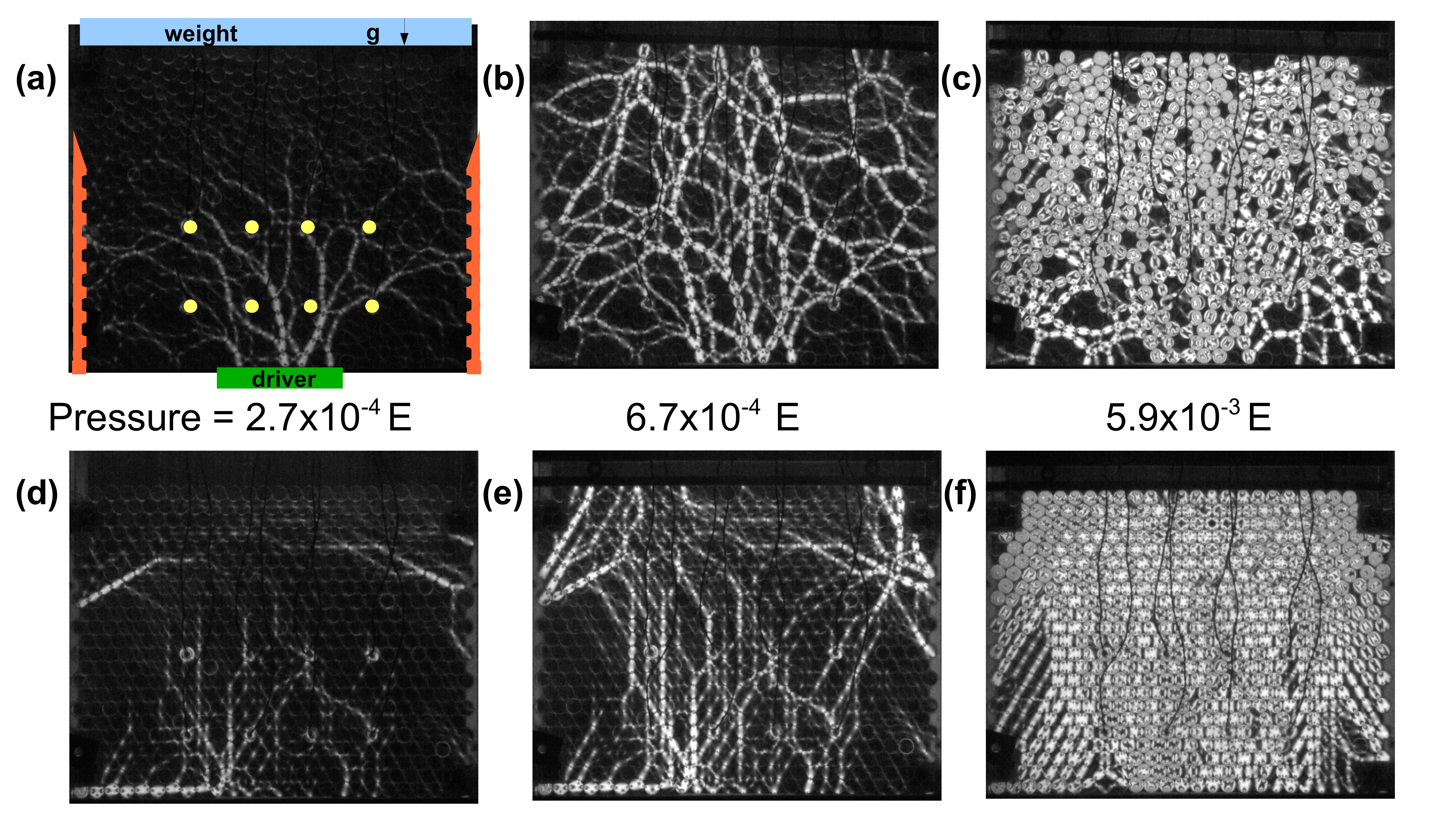}
 \caption{Images of the granular packing as a function of pressure (columns) and degree of order (rows). Image (a) highlights the location of key parts of the apparatus. The vibrating driver is located at the bottom of the granular packing, brass weights of various sizes provide the confining pressure, and rough walls restrict bulk motion. Eight particles (shown by yellow circles) contain embedded piezoelectric sensors for detecting vibrations. Images (a,b,c) show the force chains as a function of pressure for a single disordered packing, and images (d,e,f) for a single ordered packing.}
 \label{fig:exp} 
 \end{figure*}

In this paper, we will describe an alternative experimental technique for measuring vibrational modes in a frictional granular packing. Our approach is inspired by one previously used to measure the density of states via the velocity autocorrelation function (VACF) in simulations of conventional solid\cite{Dickey1969} or liquid\cite{Keyes1997} systems. In recent simulations of thermal soft spheres\cite{Wang}, this method successfully recovered the familiar scaling  $\omega^* \propto (\phi-\phi_c)^{1/2}$, where $\phi$ is the packing fraction. The measurement starts from the velocity autocorrelation function $C_v(t)$, defined as
\begin{equation}
C_v(t) \equiv \frac{\sum_{i} \langle v_i(\tau + t) \cdot v_i(\tau) \rangle_{\tau}}{\sum_{i} \langle v_i(\tau) \cdot v_i(\tau) \rangle_{\tau}}
\label{eq:Cv}
\end{equation}
where $v_i(t)$ is the velocity of particle $i$ as a function of time,  $\langle \cdot \rangle_\tau$ represents a temporal average, and the summation extends over the particles. The density of states $D(f)$ is then given by 
\begin{equation}
D(f) \equiv \int_0^{\infty} C_v(t) \cos{(2 \pi f t)} \,dt.
\label{eq:Dw}
\end{equation}

This method, which we will refer to as the VACF method, provides the thermal density of states, and its applicability to an athermal system is neither expected nor guaranteed. However, there are several criteria which, if satisfied, would give us some confidence that the analogy is a reasonable one. Ideally, we want an isotropic source of vibrations which will overcome the dissipation and provide a steady state. Second, this injected energy should partition itself equally among the degrees of freedom, and provide a thermal-like velocity distribution for each particle (corresponding to the temperature of the system). 

In our experiments, we mimic thermal vibrations using acoustic excitations at the lower boundary, and test the degree to which the rest of the criteria are satisfied. Using particle-scale measurements from piezoelectric sensors embedded in a subset of particles, we apply Eq.~\ref{eq:Cv} and \ref{eq:Dw} to quantify the resulting vibrational modes of the packing.  While the results of Eq.~\ref{eq:Dw} cannot properly be called a density of states (we shall refer to it as a density of modes), we are nonetheless able to investigate the empirical utility of such a quantity in describing the state of the material near the rigidity transition. While the VACF method provides $D(f)$, it does not allow for the visualization of the spatial modes. However, there are several key advantages of the method: we do not require optical access to the particles or need to sample all particles in the packing. These advantages would allow our method to be easily adapted to a real 3D granular materials.

\section{Experimental Setup} %========================================================================

We perform experiments in a two dimensional granular packing composed of discs cut from Vishay PSM-4, which is a photoelastic material that allows for the visualization of the internal force structure.  The granular packing has lateral dimensions of $29\times22$~cm, and is oriented vertically in order to  both minimize friction with the walls and allow the pressure to be set using brass weights. In order to minimize bulk movement of the packing, the side walls of the apparatus have been made rough (see Fig.~\ref{fig:exp}). Further details about the apparatus are available in Ref.~\cite{Owens2011}. 

In this Paper, we present experiments on both ordered and disordered particle configurations. The ordered packings are composed of monodisperse discs with diameter $a_L = 11$~mm arranged by hand into an ordered hexagonal pattern. The disordered packings are bidisperse, with an equal mixture of  $a_S = 9$~mm and $a_L = 11$~mm particles in order to suppress crystallization.  The particles have density $\rho=1.06$~g/cm$^3$, thickness $6.35$~mm, and static bulk modulus $E=4$~kPa.  We make measurements on a total of $60$ unique particle configurations, of which $30$ are disordered and $30$ are ordered.  For each of these particle configurations, we also investigate $7$ different pressures ranging from $2.7\times10^{-4}$ to $5.9\times10^{-3}$~$E$, where the lowest pressure is set by the weight of the particles themselves.   

An electromagnetic driver (MB Dynamics PM50A) is attached to a driving platform of width $8.5$~cm located at the bottom of the packing. Because a granular material is dissipative (due to both friction and viscoelasticity), continuous driving is necessary to maintain a steady state. We mimic thermal vibrations by subjecting the platform to oscillations which have a flat velocity spectrum. Observations of a piezoelectric sensor mounted on the driver platform indicate that the system (apparatus plus the granular material) has several strong mechanical resonances. We are able to compensate for these resonances up to $3$~kHz and maintain a flat driver response down to approximately $300$~Hz.  Below $300$~Hz, we approximate the driver response $\tilde{A}(f)$ by a third degree polynomial for the disordered packing and a power law for the ordered packing; these empirical measurements will later be used to correct the sensor responses.

Vibrations are recorded from eight particles which contain a piezoelectric sensor at their center. The particles are arranged in two rows, the first one $4.5$~cm from the bottom wall and the second at $10$~cm (see Fig.~\ref{fig:exp}). Piezoelectric materials are sensitive to a single direction of stress, and therefore only measure the component of acceleration along that axis. During setup, individual sensors are placed in a variety of orientations. This ensures that the ensemble of sensors collects data from many different directions, both multiply-scattered waves and some which arrive directly from the driver.

\begin{figure}
  \includegraphics[width=.9\linewidth]{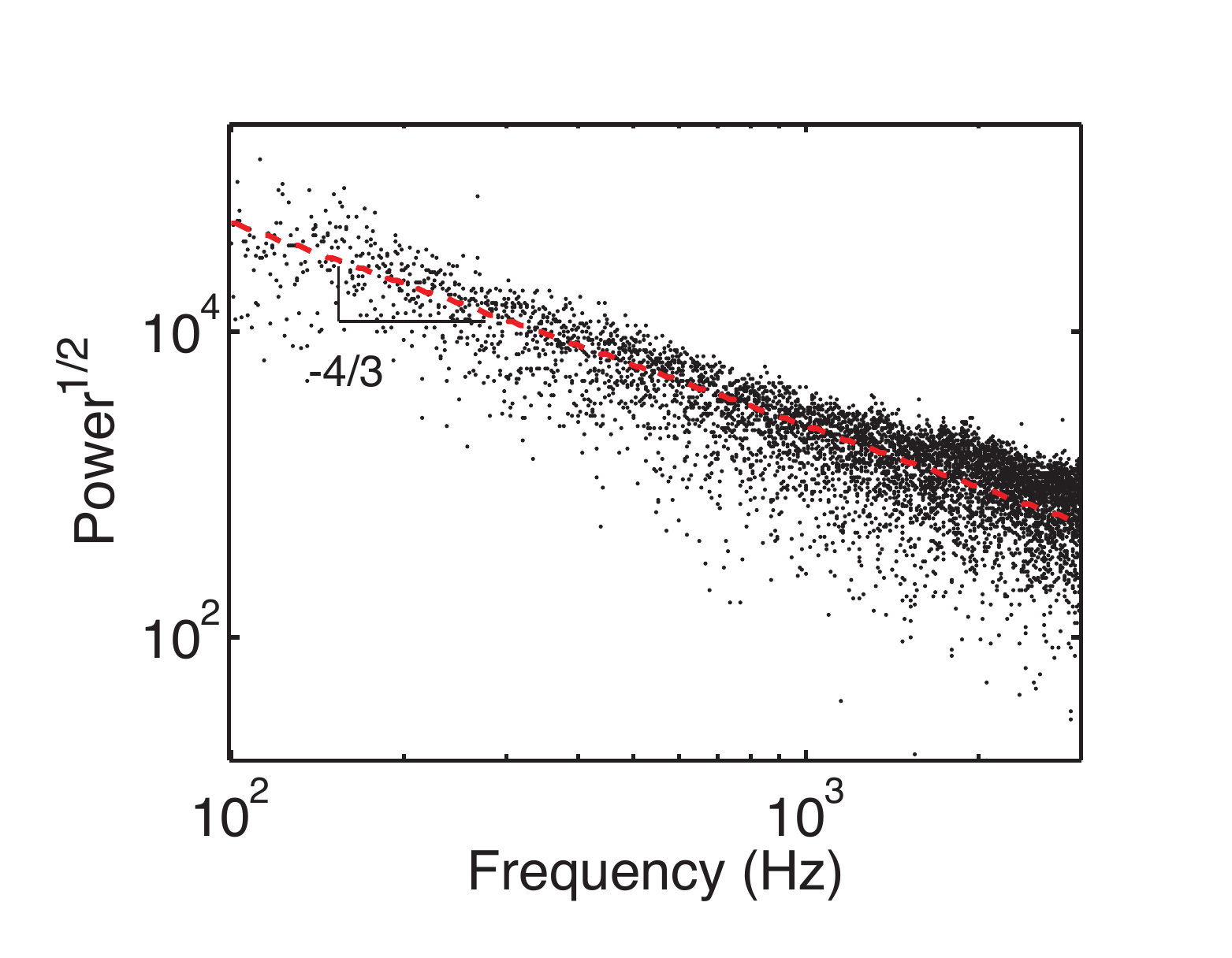}
  \centering
 \caption{Velocity spectrum measured in a continuous sheet of PSM-4 material, showing $f^{-4/3}$ decay.}
\label{fig:bulk} 
\end{figure} 

Piezoelectric sensors produce a voltage $V(t)$ proportional to the stress on the sensor, and therefore also proportional to its instantaneous acceleration. Each sensor is sampled for a duration of $2$~sec at a sampling rate of $100$~kHz. By integrating  $v'(t) = \int V(t)dt$, we obtain a velocity measurement $v'$, which must still be corrected by deconvolving both the viscoelastic response of the bulk material and the low-frequency driver response $\tilde{A}(f)$. As the measured voltages are difficult to calibrate, all velocities will be reported in arbitrary, but consistent, units. 

In order to factor out the viscoelasticity of the photoelastic material, we measure the response of a continuous sheet of PSM-4 embedded with identical piezoelectric sensors, subject to the same driving input as the granular experiments. The result of this test is shown in Fig.~\ref{fig:bulk}. We observe that the velocity spectrum exhibits a $f^{-4/3}$ decay.  Strong, frequency dependent, damping is expected for viscoelastic materials\cite{visc}.

In Fourier space (denoted by $\sim$), we combine these measurements to obtain $\tilde{v} = \tilde{v}' \, f^{4/3} / \tilde{A}$.  We additionally apply a band-pass filter with a low-frequency cutoff of $f=75$Hz and a high-frequency cutoff of $f=3$~kHz, as these regimes are dominated by electrical noise and the mechanical response of the apparatus, respectively. Finally, we perform an inverse transform and use the resulting $v(t)$ for the remainder of the analysis.

By using birefringent particles, we can measure the particle positions via a Hough transform\cite{Hough} and the vector contact forces via a nonlinear fit to  the photoelastic fringes visible through crossed circular polarizers\cite{Puckett2012c, peDiscSolve}. We make these measurements on particles located in the same region as the sensor particles. We exclude rattlers (particles with less than two detectable neighbors) from the analysis.

In order to make comparisons with computer simulations, it is helpful to non-dimensionalize our measurements. Following convention\cite{Silbert2005}, we take our frequency unit to be $f_0=\frac{1}{2\pi}\sqrt{\frac{V_0}{m a^2}}$, where $m$ is the mass of the particles, $a$ is the mean particle diameter, and $V_0$ comes from the harmonic particle contact law used in the simulation. 
To estimate $V_0$ for our particles, we compare the force law that corresponds to the interaction potential ($F = \frac{2V_0}{a^2}\delta$, where  $\delta$ is the Hertzian overlap) to the force law for ideal Hertzian discs\cite{johnson} ($F=\frac{\pi EL}{8(1-\nu^2)} \delta$,  where $L$ is the particle thickness and $\nu$ is the Poisson ratio). Matching the coefficients, we find that $f_0(f) = \sqrt{\frac{\, L \, E(f)}{64 \pi m (1 - \nu^2)}}$. This treatment neglects the observation that our particles deviate from the linear Hertzian law,\cite{Owens2011} and instead have a force law $F\propto\delta^{5/4}$.  Due to the viscoelasticity\cite{Owens2011} of our particles, the modulus is frequency-dependent, with modulus $E(f) \propto f^{1/2}$. Therefore, $f_0$ is itself frequency-dependent, and we report frequencies as the ratio $\frac{f}{f_0(f)}$ in order to make comparisons with simulations. The range of accessible non-dimensional frequencies corresponds to $0.06$ to $0.8$.

 \begin{figure*}
  \centering 
\includegraphics[width=\linewidth]{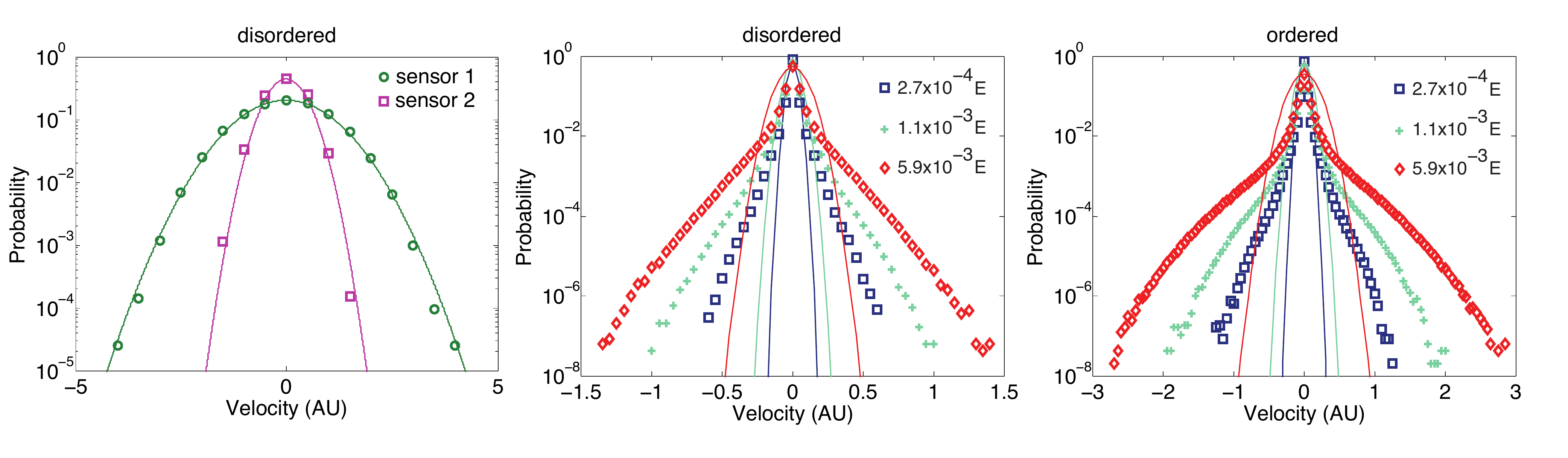}
\centering
 \caption{(a) Two velocity distributions ${\cal P}_i(v)$, for two sensored particles in the first row of the same disordered packing, with pressure $P = 2.2 \times 10^{-3}\,E$. Ensemble velocity distributions ${\cal P}(v)$ of all sensored particles, rows, and experiments in  (b) disordered packings and (c) ordered packings. Gaussian distributions with the same standard deviation as the data are shown as solid lines on all plots.}
 \label{thermal} 
 \end{figure*}

\begin{figure}
  \includegraphics[width=.8\linewidth]{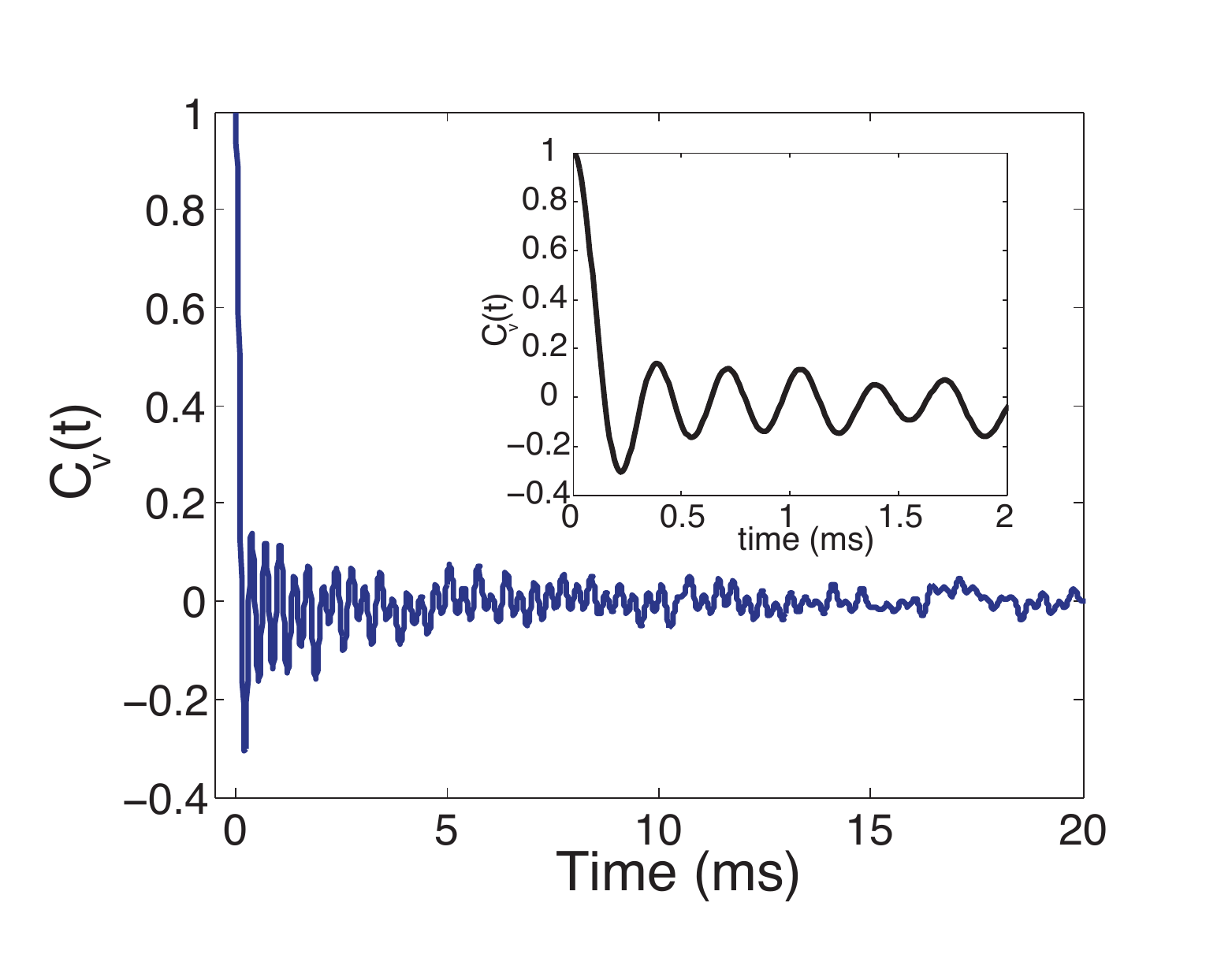}
  \centering 
 \caption{Example velocity autocorrelation function (from Eq.~\ref{eq:Cv}), in a disordered packing with $P = 5.9x10^{-3} \, E$. Correlations decay within approximately $20$~ms for all experimental runs.}
 \label{fig:Cv} 
 \end{figure}

\section{Results} %================================================================================================
 
\subsection{Thermal analogy}

First, we characterize the extent to which the analogy to thermal systems is reasonable. We anisotropically inject energy from a single boundary of the granular packing for $2.8$~sec, of which the middle $2$~sec is analyzed. The typical time for $V(t)$ measurements to either reach steady state or to decay when driving ceases is on the order of $1$~ms.

In a thermal solid, particles should have a Gaussian velocity distribution, where the standard deviation of the distribution is set by the temperature of the material. For each sensored particle, in each experiment, we measure the velocity distributions ${\cal P}_i(v)$ (see Fig.~\ref{thermal}), and find that these  individual distributions are well-described by Gaussian distributions; the average kurtosis of the distributions is $3.0 \pm 0.07$. However, different particles, even from the same row of the same experiment, have different widths, depending on the local environment. From the measured vector contact forces on each sensored particle, we observe that the width of a particle's velocity distribution increases with the local pressure from its neighbors. The trend is consistent with what was previously observed\cite{Owens2011} for measurements of sound wave amplitudes, and can be attributed to an increased transmission of energy through contacts with larger contact area.

As a result of these spatial heterogeneities, the velocity distribution ${\cal P}(v)$ measured over all 30 experiments at the same $P$ is not a Gaussian. Instead, as shown in Fig.~\ref{thermal}bc, the tails are significantly broader; this holds for both the disordered and ordered packings. In addition, the ensemble distributions have a width which increases with increasing pressure. This means that the ensemble does not satisfy equipartition, and we must therefore proceed with caution.

\subsection{Determining $D(f)$}

\begin{figure*}
  \includegraphics[width=1\linewidth]{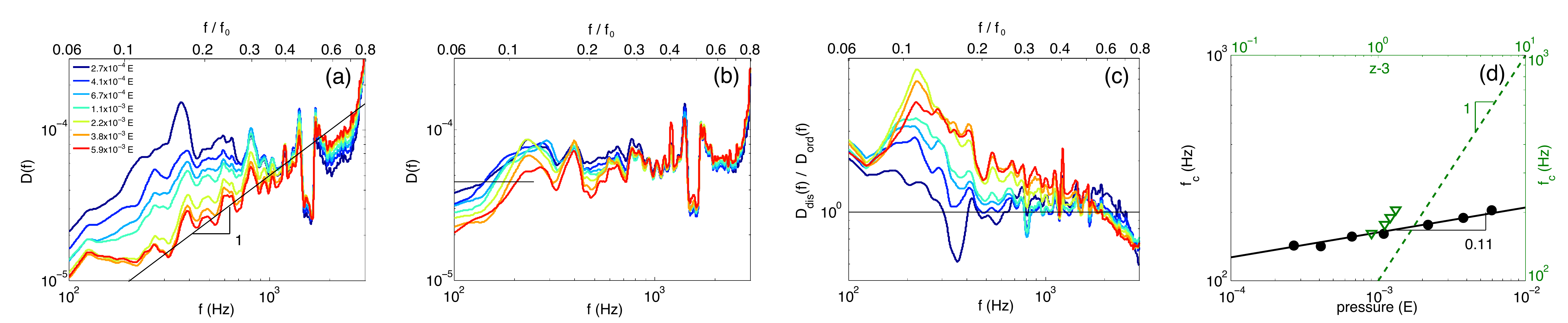}
 \caption{Measured density of modes $D(f)$ as a function of confining pressure $P$ for (a) ordered and (b) disordered packings. Line color corresponds to $P$ as shown in the legend. The frequencies are reported in both experimental frequency units (bottom, Hz) and simulation units (top, $f/f_0(f)$). In (a), the solid black line provides a comparison to Debye scaling for 2D systems. In (b), the solid line shows the threshold specifying the cutoff frequency $f_c$. (c) Ratio of the ordered and disordered $D(f)$. (d) Scaling of $f_c$ as a function of $P$ (bottom axis) and $z-3$ (top axis).}
 \label{fig:DOM} 
 \end{figure*}

In order to utilize the VACF method, we would ideally use all particles in a packing when calculating the velocity autocorrelation $C_v(t)$ from Eq.~\ref{eq:Cv}. Instead, we average over eight particles within 30 experiments at the same pressure and same order/disorder, and obtain a long time-average for each.  An example $C_v(t)$ is shown in Fig.~\ref{fig:Cv}; note that the decay time is much shorter ($20$~ms) than the $2$~sec measurement duration, providing many decorrelated measurements during each experiment. Using the real part of the spectrum of $C_v(t)$, we calculate the density of modes $D(f)$ via Eq.~\ref{eq:Dw}. There are strong peaks at multiples of $60$~Hz due to electrical noise which we filter out as a last step after calculating $D(f)$.

We calculate $D(f)$ for each of seven confining pressures $P$, applied to 30 ordered and 30 disordered packings.
The results are shown in Fig.~\ref{fig:DOM}ab. We observe Debye scaling ($f^{d-1}$, for dimension $d=2$) for large $P$ in the ordered packings, which is expected since these packings are the most uniform in the coordination number $z$ and and contact forces. Debye scaling is not observed in the disordered experiments. 

In order to compare the ordered and disordered packings, it is helpful to consider the ratio $D_{dis}/D_{ord}$. In Fig.~\ref{fig:DOM}c, we plot this ratio and thereby eliminate some of the remaining electronic/apparatus resonances: note that the sharp decrease near $1500$~Hz, as well as number of smaller features, are removed. We observe from this ratio that at all values of $P$, disordered packings have more low-frequency modes than do the ordered packings.

As a function of decreasing $P$, both ordered and disordered packings exhibit a growing number of low-frequency modes. In the case of the ordered packings, it is important to note that at low $P$, disorder is still present in both the contact network ($z \le 6$ for many of the particles) and the force chain network (see Fig.~\ref{fig:exp}). For the disordered packings, we quantify this decrease in the number of low-frequency modes by calculating a critical frequency $f_c$ at which $D(f)$ falls below a threshold (see Fig.~\ref{fig:DOM}b). 

Fig.~\ref{fig:DOM}d shows how $f_c$ depends on the confining pressure $P$; we fit a power law $f_c \propto P^{0.11 \pm 0.05}$, where the mean and uncertainty is determined from a bootstrapping method that repeats the analysis for subsets of the data, randomly resampled with replacement.  For a frictionless packing with Hertzian exponent $\alpha = 5/4$, $f_c$ would be expected to scale as $f_c \propto P^{\beta}$, with an exponent $\beta = 2/5$; our observations show a considerably flatter trend. This can be understood in light of simulations of frictional particles\cite{Somfai2007}, where power-law scaling was only observed in the large-friction limit. For experimentally-reasonable friction coefficients ($\mu = 0.2$ to $0.8$), no simple ($f_c$,$P$) scaling was observed. However, they did observe a friction-independent scaling relationship  $\omega_c \propto (z-3)$ (where $z_c = 3$ is the isostatic value for a frictional packing); \citet{Henkes2010} observed a similar scaling relation. In our experimental results, good estimates of $z-3$ are only available for $P \gtrsim 10^{-3} E$; this data is plotted in Fig.~\ref{fig:DOM}d for comparison and shows the expected steeper trend. The range of measured values is unfortunately too small to draw a more quantitative conclusion.

\section{Discussion}

In calculating a density of modes $D(f)$ for our athermal, frictional, viscoelastic, granular packings, we mimic thermalization using acoustic excitation. We find that while some thermal-like conditions are satisfied --  we can achieve an isotropic, steady state through continuous driving and measuring vibrations resulting from multiply-scattered waves -- the velocity distributions are fundamentally non-thermal-like.  While individual particles have Gaussian velocity distributions, each one is sufficiently different that the ensemble distribution does not. Notably, experiments by \citet{Brito2010} also found that displacement distributions were individually Gaussian, but with widths which varied from particle to particle. Through the use of photoelastic particles, we explain these variations as arising from spatial heterogeneities in the force chains. In spite of these non-thermal features, the VACF method provides a means to calculate $D(f)$, and the observation of Debye scaling for the most uniform (well-compressed, ordered) packings suggest that the analogy is justified.

The VACF method differs significantly from what has been utilized in prior experiments\cite{Chen2010,Ghosh2010a,Brito2010}, in which particle trajectories are used to construct a covariance matrix\cite{Henkes2012}. The covariance matrix method works well in systems where there is optical access to accurate position information for a large number of particles, and has the benefit of providing a means to visualize the spatial modes at each frequency. However, real granular packings do not typically provide the necessary optical access for such methods to be feasible. In addition, the quality of displacement measurements depends strongly on the frame rate of the camera, which is often at odds with the high spatial resolution needed for accurate particle-tracking. In contrast, our application of the VACF method does not require optical access to the packing, and instead relies on only a small number of piezoelectric sensors which readily provide high temporal resolution. While we have used photoelastic particles in order to visualize the internal stresses, the VACF method does not require such a packing, and would be applicable for use in ordinary, non-circular, three-dimensional granular packings.

A useful application of a granular $D(f)$ would be to measure how far a packing is from $P=0$, $\phi_c$, or $z_c$. While we observe only a flat trend ($f_c \propto P^{0.11}$), there is evidence for a stronger dependence on the coordination number. This is very encouraging, since the presence/absence of contacts has been extremely difficult to measure accurately in experiments. Further investigations should explore this scaling relation more quantitatively.

\section{Conclusion}
 
Our experiments use a novel technique to apply thermal methods to a real, athermal, granular packing through the use of acoustic excitation.  We have probed the mode structure of a granular packing by acoustically mimicking thermal vibrations and calculated a density of modes $D(f)$ using the spectrum of the velocity autocorrelation function.  Our granular packings deviate in many important ways from the ideal ones used in simulations since our particles are soft, frictional, and viscoelastic; furthermore, there is no equipartition of vibration amplitudes among the particles. Nonetheless, we can recover many important features such as Debye scaling at high pressure in the ordered packings and an excess number of low-frequency modes compared to Debye scaling in the disordered packings. These low-frequency modes can be used to differentiate the pressure state of the packing as it is seen that a crossover frequency $f_c$ varies with both pressure and coordination number. We have thereby been able to use a small number of particle scale measurements to see system-level properties, a technique which could be extended to granular materials where there is no optical access.   
 
\section{Acknowledgments}

We thank Corey O'Hern, Carl Schreck, Silke Henkes, Olivier Dauchot, and Thomas Owens for helpful discussions, James Puckett for sharing his photoelastic methods,  the West Virginia High Technology Consortium Foundation for donation of the shaker and isolation table used in experiments.  This work was supported by NSF DMR-0644743 and DMR-1206808.


%\balance


%\footnotesize{
%\bibliographystyle{rsc} %the RSC's .bst file
%\bibliography{eto2,ked} %your .bib file
%}

\providecommand*{\mcitethebibliography}{\thebibliography}
\csname @ifundefined\endcsname{endmcitethebibliography}
{\let\endmcitethebibliography\endthebibliography}{}

\end{document}